# Bending-induced isostructural transitions in ultrathin layers of van der Waals ferrielectrics

Anna N. Morozovska[1], Eugene A. Eliseev[2], Yongtao Liu[3], Kyle P. Kelley[3], Ayana Ghosh[3], Ying Liu[4], Jinyuan Yao[4], Nicholas V. Morozovsky[1], Andrei L Kholkin[5*], Yulian M. Vysochanskii[6†], and Sergei V. Kalinin[7‡]

[1] Institute of Physics, National Academy of Sciences of Ukraine,
46, Prospekt Nauky, 03028 Kyiv, Ukraine
[2] Institute for Problems of Materials Science, National Academy of Sciences of Ukraine,
[3]Computational Sciences and Engineering Division, Oak Ridge National Laboratory, Oak Ridge, Tennessee, 37831
[4]Department of Physics and Materials Research Institute, Pennsylvania State University, University Park, Pennsylvania 16802, United States
[5] Department of Physics & CICECO – Aveiro Institute of Materials, Campus Universitario de Santiago, 3810-193 Aveiro, Portugal
3, Krjijanovskogo, 03142 Kyiv, Ukraine
[6] Institute of Solid-State Physics and Chemistry, Uzhhorod University,
88000 Uzhhorod, Ukraine
[7] Department of Materials Science and Engineering, University of Tennessee,
Knoxville, TN, 37996, USA

## Abstract

Using Landau-Ginzburg-Devonshire (LGD) phenomenological approach we analyze the bending-induced re-distribution of electric polarization and field, elastic stresses and strains inside ultrathin layers of van der Waals ferrielectrics. We consider a $CuInP_2S_6$ (CIPS) thin layer with fixed edges and suspended central part, the bending of which is induced by external forces. The unique aspect of CIPS is the existence of two ferrielectric states, FI1 and FI2, corresponding to big and small polarization values, which arise due to the specific four-well potential of the eighth-order LGD functional. When the CIPS layer is flat, the single-domain FI1 state is stable in the central part of the layer, and the FI2 states are stable near the fixed edges. With an increase of the layer bending below the critical value, the sizes of the FI2 states near the fixed edges decreases, and the size of the FI1 region increases. When the bending exceeds the critical value, the edge FI2 states disappear being substituted by the FI1 state, but they appear abruptly near the inflection regions and expand as the bending increases. The bending-induced isostructural FI1-FI2 transition is specific for the bended van der Waals ferrielectrics described by the eighth (or higher) order LGD functional with consideration of linear and nonlinear electrostriction couplings. The isostructural transition, which is revealed in the vicinity of room temperature, can significantly reduce the coercive voltage of ferroelectric polarization reversal in CIPS nanoflakes, allowing for the curvature-engineering control of various flexible nanodevices.

[*] Corresponding author, e-mail: kholkin@ua.pt

[†] Corresponding author, e-mail: vysochanskii@gmail.com

[‡] corresponding author, e-mail: sergei2@utk.edu

## I. INTRODUCTION

Despite the rapid growth of research efforts targeting lead-free low-dimensional ferroelectric materials, their polar and piezoelectric properties, and phase diagrams remain relatively unexplored. At the same time, these properties define their potential applicability for nanoelectronics, piezo-harvesting, and energy storage [1, 2]. Among multiple families of the 2D ferroelectrics, much attention has been attracted to lead-free low-dimensional van der Waals (**vdW**) ferroics such as $CuInP_2(S_xSe_{1-x})_6$ [3, 4] due to robust ferroelectricity, ferrielectricity, and antiferroelectricity in their monolayers, thin films and nanoflakes. However, the polar, piezoelectric and dielectric properties of these low-dimensional vdW ferroics are poorly known and often anomalous in comparison with well-studied oxide perovskite ferroelectrics [5, 6, 7]. For example, the out-of-plane spontaneous polarization increases strongly for compressive strains and decreases for tensile strains of $CuInP_2(S_xSe_{1-x})_6$ ultrathin layers [8, 9, 10, 11], whereas the out-of-plane polarization increases for tensile strains and vanishes for compressive strains in classical ferroelectrics [12]. Possible reasons of the controversial and anomalous behaviors are size effects [13], giant negative electrostriction [14, 15] and flexoelectric [16, 17] couplings. In addition, the existence of more than two potential wells [18, 19, 20], which are responsible for strain-tunable multiple polar states, plays a very important role in the physical properties of the ferrielectric $CuInP_2(S,Se)_6$, compared to many other uniaxial ferroelectrics-antiferroelectrics.

Ferrielectricity is closely related to an antiferroelectric order, but with a switchable spontaneous polarization created by two sublattices with antiparallel and different in magnitude spontaneous dipole moments [21]. In accordance with this definition and experimental results [22, 23, 24], $CuInP_2(S_xSe_{1-x})_6$ is a ferrielectric material, because it has two polar sublattices, Cu and In, which electric dipole moments are antiparallel and different in magnitude in the absence of external electric field and applied stresses or strains.

Despite a gamut of fascinating experimental results [13 - 17], there are few theoretical studies of strain-induced phenomena in bended low-dimensional vdW ferroelectrics, such as ultrathin $CuInP_2S_6$ layers. In particular, Yang et al [17] explored the ferroelectric domain switching in an ultrathin $CuInP_2S_6$ nanoflake by applying an electric field and mechanical loading in atomic force microscopy, revealed the loading-induced polarization reversal and explained the observed decrease in the coercive voltage of domain switching by the Landau-Ginzburg-Devonshire model. Yang et al noted that the strain-induced reduction of the coercive voltage is highly required for the $CuInP_2S_6$ nanoflakes applications, as e.g., memory cells.

Using the Landau-Ginzburg-Devonshire (**LGD**) approach and the Finite Element Modeling (**FEM**), here we analyze the bending-induced re-distribution of electric polarization and field, elastic stresses and strains inside ultrathin layers (10 nm or less) of vdW ferroelectrics. As an example, we consider the ultrathin $CuInP_2S_6$ (**CIPS**) layer with fixed edges and suspended central part, which bending is controlled by external forces, such as attraction force of a rough substrate. The bending influences the so-called low-polarization states in CIPS, which have the spontaneous polarization $P_S \approx (3 – 4)$ μC/cm$^2$ and are studied in this work. The original part of this work contains the physical description of the problem (**Section II**) and analysis the bending-induced transitions of a polarization in the CIPS layer (**Section III**). **Section IV** summarizes the obtained results and potential outcomes. **Supplementary Materials** elaborate on a mathematical formulation of the problem and contain a table of material parameters.

## II. BASIC EQUATIONS AND APPROXIMATIONS

### B. Finite Element Modeling details

The electric polarization and field, elastic stresses and strains of a bended CIPS layer are studied by the FEM. We consider a thin layer of thickness $h$, which models a CIPS nanoflake, covered by ideally-conductive and elastically soft electrodes or/and screening charge layers, which do not influence the layer bending [see **Fig. 1(a)**]. The electric contact between the layer and conductive electrodes is regarded to be ideal. Initially, the Z-axis is parallel to the polar axis of the flat layer, and its polarization $P_3 \uparrow\uparrow Z$. Since the bending changes the shape of the layer, we introduce the local coordinate frame linked with the instant position of crystallographic axes inside the bended layer, and distinguish the normal ("$n$") and tangential ("$t$") directions in the local coordinate frame $\{x_t, x_n\}$. We consider the layer with fixed edges, and, to imitate the bending occurring due to the attraction force (e.g., adhesion or/and disperse London forces), we applied a spatially-distributed force in the normal direction Z. The force, which density has a parabolic spatial distribution, is maximal in the center of the layer and zero at the fixed edges. The force density, $\sigma_f$, is measured in the pressure units, $[\sigma_f] = \frac{N}{m^2}$ =Pa.

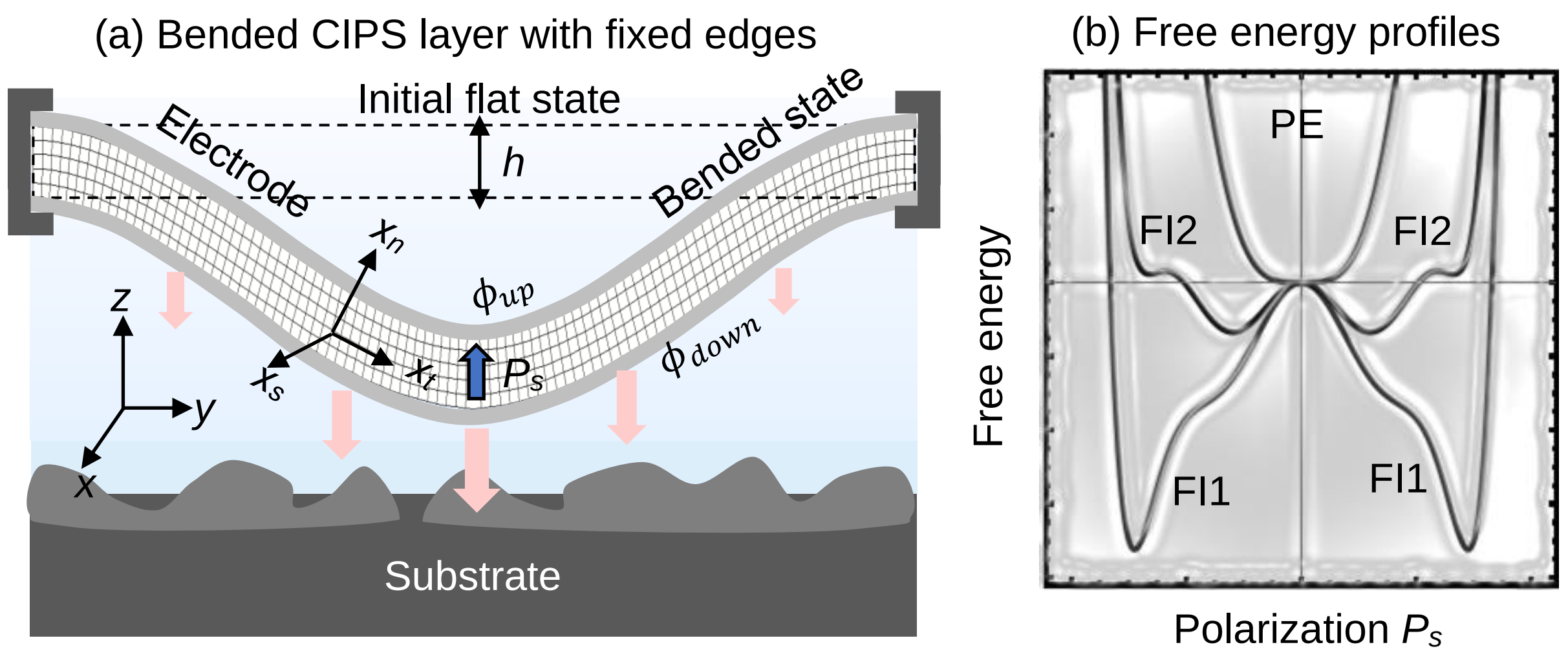

**FIGURE 1**. **(a)** Bended CIPS layer with fixed edges, its "up" and "down" surfaces are covered by ideally-conductive and elastically soft ultrathin electrodes. Thick pink arrows illustrate the attractive force, which bends the CIPS layer. **(b)** Schematics of the CIPS free energy relief in the paraelectric (PE) phase (the single-well curve), ferrielectric state FI2 (the four-well curve) and ferrielectric state FI1 (the double-well curve with or without inflection points). The part (b) is adapted from Ref.[25].

We performed the FEM in COMSOL@MultiPhysics software, using electrostatics, general math (PDE toolbox), and solid mechanics modules for fixed elastics boundary conditions. The size of the computational region is 200 nm ×2 nm×10 nm, and periodic boundary conditions are applied in the smallest direction Y. The minimal size of a cubic mesh element is equal to the 0.5 nm, and the maximal size is 2 nm. The mesh size increase results in minor changes in the physical quantities, such that the spatial distribution of each of them becomes less smooth, but all significant details remain visible. The FEM was performed in the temperature range (250 – 350) K, and the special attention was paid to the room temperature and its vicinity, since the most interesting effects occur near the Curie temperature of CIPS, $T_C$ = 292.67 K. In FEM, we use the constitutive equations for relevant order parameter fields based on the LGD approach along with electrostatic equations and elasticity theory.

### B. Landau-Ginsburg-Devonshire approach

The LGD free energy functional $G$ of a uniaxial ferrielectric CIPS includes a Landau energy – an expansion on 2-4-6-8 powers of the polarization components $P_i$, $G_{Landau}$; a polarization gradient energy, $G_{grad}$; an electrostatic energy, $G_{elect}$; elastic and electrostriction energies, $G_{el+els}$; a flexoelectric energy, $G_{flexo}$; and a surface energy, $G_S$. Therefore, $G$ has the form [26]:

$$G = G_{Landau} + G_{grad} + G_{elect} + G_{elas} + G_{flexo} + G_S, \quad (1a)$$

$$G_{Landau} = \int_{V_f} d^3r \left(\frac{\alpha}{2}P_n^2 + \frac{\beta}{4}P_n^4 + \frac{\gamma}{6}P_n^6 + \frac{\delta}{8}P_n^8\right), \quad (1b)$$

$$G_{grad} = \int_{V_C} d^3r \frac{g_{ijkl}}{2}\frac{\partial P_i}{\partial x_k}\frac{\partial P_j}{\partial x_l}, \quad (1c)$$

$$G_{elect} = -\int_{V_f} d^3r \left(P_i E_i + \frac{\varepsilon_0 \varepsilon_b}{2} E_i E_i\right), \tag{1d}$$

$$G_{el+els} = -\int_{V_f} d^3r \left(\frac{s_{ijkl}}{2}\sigma_{ij}\sigma_{kl} + \frac{Q_{ijkl}}{2}\sigma_{ij}P_k P_l + \frac{Z_{ijkl}}{4}\sigma_{ij}P_k^2 P_l^2 + \frac{W_{ijklmq}}{2}\sigma_{ij}\sigma_{kl}P_m P_q\right), \tag{1e}$$

$$G_{flexo} = -\int_{V_f} d^3r \frac{F_{ijkl}}{2}\left(\sigma_{ij}\frac{\partial P_k}{\partial x_l} - P_k\frac{\partial \sigma_{ij}}{\partial x_l}\right), \tag{1f}$$

$$G_S = \frac{1}{2}\int_S d^2r\, a_{nn}^{(S)} P_n^2. \tag{1g}$$

The integration is performed over the volume $V_f$ of the nanoflake. Einstein summation over repeated indexes is used in Eqs.(1c)-(1f), and the subscripts $i, j, k, l, m$ and $q$ take values $\{t, n\}$. The coefficient α in the Landau energy (1b) linearly depends on temperature $T$, $\alpha(T) = \alpha_T(T - T_C)$, where $\alpha_T$ is the inverse Curie-Weiss constant and $T_C$ is the ferroelectric Curie temperature. Coefficients β, γ and $\delta$ in Eqs.(1b) are regarded as temperature-independent. The coefficients $g_{ijkl}$ in the gradient energy (1c) are positively defined. The electrostatic energy (1d) is written in the form suggested by Marvan and Fousek [27]; $E_i$ is the sum of applied and depolarization field, parameters $\varepsilon_0$ and $\varepsilon_b$ are the universal dielectric constant and the relative background dielectric permittivity of the ferrielectric [28, 29], respectively. In Eq.(1e), $\sigma_{ij}$ is the stress tensor, $s_{ijkl}$ is the elastic compliances tensor, $Q_{ijkl}$, $Z_{ijkl}$, and $W_{ijklmn}$ are the linear and nonlinear electrostriction tensor components, respectively. The flexoelectric stress tensor $F_{ijkl}$ in Eq.(1f) is determined by the microscopic properties of the material [30].

The CIPS parameters used in our calculations are listed in **Table SI**. For these parameters, in dependence on the temperature and applied stress or strain, the CIPS free energy relief corresponds either to the paraelectric (**PE**) phase being the single-well, or to ferrielectric state 2 (**FI2**) being the four-well, or to the ferrielectric state 1 (**FI1**) being the double-well with or without inflection points [see the schematics in **Fig. 1(b)**].

Allowing for Landau-Khalatnikov relaxation mechanism [31], variation of the free energy (1) with respect to the ferroelectric polarization gives the following equation:

$$\Gamma\frac{\partial P_n}{\partial \tau} + \left[\alpha - \sigma_{ij}\left(Q_{ijnn} + W_{ijklnn}\sigma_{kl}\right)\right]P_n + \left(\beta - Z_{ijnn}\sigma_{ij}\right)P_n^3 + \gamma P_n^5 + \delta P_n^7 - \tilde{g}_{nnkl}\frac{\partial^2 P_n}{\partial x_k \partial x_l} = E_n + F_{ijkn}\frac{\partial \sigma_{ij}}{\partial x_k}. \tag{2a}$$

Here the normal components of polarization and field, $P_n$ and $E_n$, are directed along the local polar axis $x_n$, $\Gamma$ is the Khalatnikov kinetic coefficient, and $\tau$ is the time. The components of the tensor $\tilde{g}_{nnkl}$ are proportional to the convolution of the gradient tensor components, $g_{ijkl}$, and Christoffel symbols [32] originated from the layer curvature. Note that the difference between $\tilde{g}_{nnkl}$ and $g_{ijkl}$ is negligible for small curvatures, and strongly increases with the curvature increase.

The boundary condition for the $P_n$ at the curved surface S is "natural":

$$\tilde{g}_{nnlk} e_k \left.\frac{\partial P_n}{\partial x_l}\right|_S = 0, \tag{2b}$$

where $e_k$ is the outer normal to the surface.

The electric field, $E_i = -\frac{\partial \phi}{\partial x_i}$, is expressed via the potential $\phi$, which satisfies the Poisson equation inside the ferrielectric,

$$\varepsilon_0 \varepsilon_b \frac{\partial^2 \phi}{\partial x_k \partial x_k} = \frac{\partial P_i}{\partial x_i}. \tag{3a}$$

Equation (3a) is supplemented by the conditions of zero potentials at the “up” and “down” CIPS surfaces, covered by electrodes:

$$\phi_{up}|_S = 0, \qquad \phi_{down}|_S = 0. \tag{3b}$$

The electric charge is zero at the fixed edges.

Modified Hooke’s law, relating elastic strains $u_{ij}$ and stresses $\sigma_{kl}$, is obtained from the relation $u_{ij} = -\partial g_{LGD}/\partial \sigma_{ij}$:

$$u_{ij} = s_{ijkl}\sigma_{kl} + Q_{ijkl}P_k P_l + Z_{ijkl}P_k^2 P_l^2 + W_{ijklmq}\sigma_{kl}P_m P_q + F_{ijkl}\frac{\partial P_k}{\partial x_l}. \tag{4}$$

Remarkably, that the multiple bending mechanisms suggest that the same bending profile can correspond to different forces if other boundary condition are not defined. Below we analyze the polar state of the CIPS layer with fixed edges and a suspended central part, which bending is controlled by external forces, such as the attraction force by a rough substrate [see **Fig. 2(a)**]. The layer is compressed above the neutral surface and stretched below the surface in the bended central region [see **Fig. 2(b)**]. At the same time, the layer is locally flat in the inflection regions, as shown in **Fig. 2(c)**.

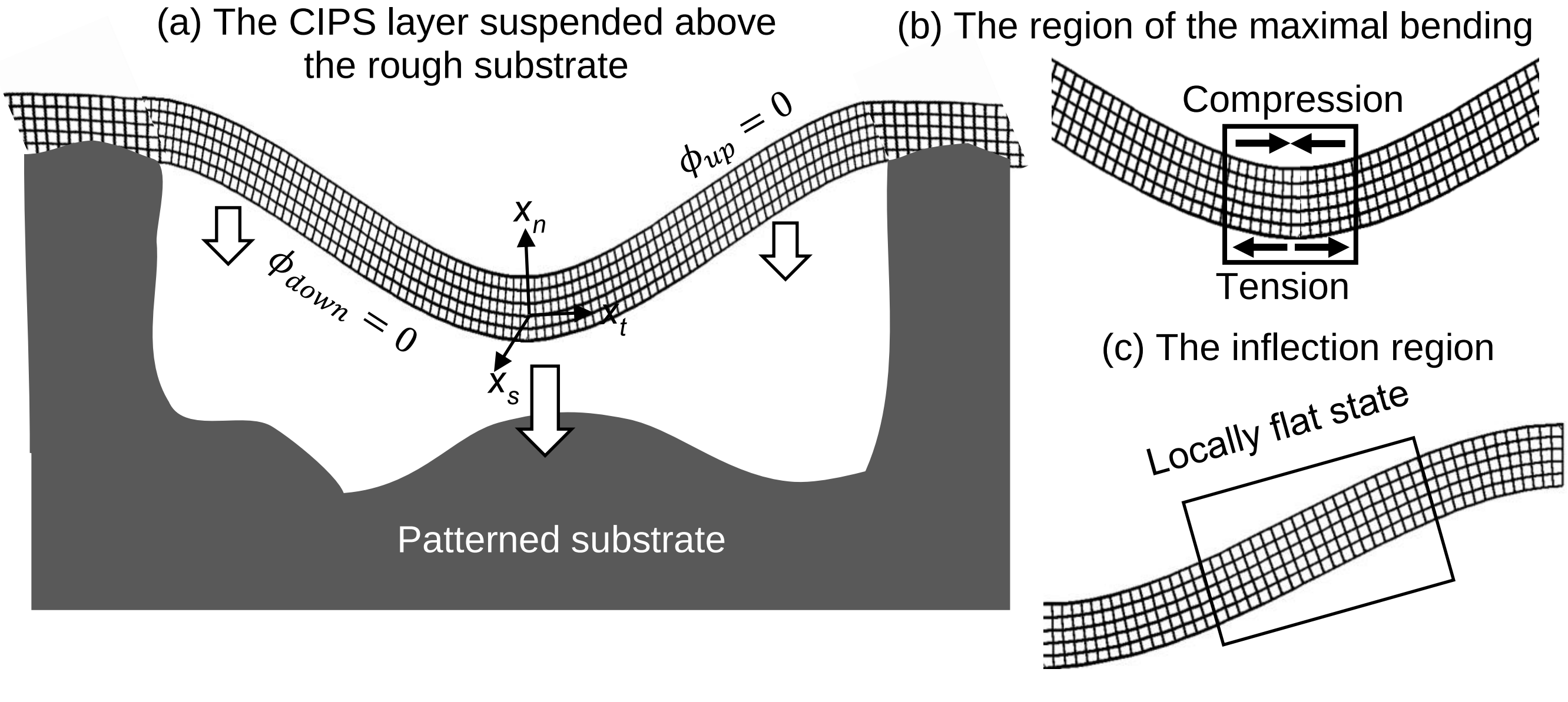

**FIGURE 2**. **(a)** A CIPS layer, which edges are clamped by a rough substrate and become fixed. Thick arrows schematically show the attraction to the substrate, which bends the suspended part of the layer. **(b)** The layer is compressed above the neutral surface and stretched below the surface in the region of a maximal bending. **(c)** The layer is locally flat in the inflection regions.

## III. RESULTS AND DISCUSSION

### A. Bended CIPS layer with fixed edges: FEM results

Since the bending changes the shape of the CIPS layer, FEM results in **Figs. 3-4** are presented in the local coordinate frame linked with the instant position of crystallographic axes inside the bended layer. The local normal axis $x_n$ coincides with $z$, and the tangential axis $x_t$ coincides with $x$ in the flat CIPS layer with fixed edges. With this, the distribution of the normal and tangential polarization components, $P_n$ and $P_t$, inside the flat layer is shown in the top row in **Figs. 3(a)** and **3(b)**, respectively. Since the top and bottom surfaces of the layer are perfectly screened, the single-domain FI1 state with the polarization $P_n \approx 4$ μC/cm$^2$ is stable in the central part of the flat layer. The FI2 states with much lower polarization $P_n \approx 0.4$ μC/cm$^2$ are stable at a distance of about 25 nm near the fixed edges. The boundary between the FI1 and FI2 regions is relatively sharp (≈ 2.5 nm) because it is uncharged, and thus uncompensated bound charges are absent at the boundary and inside the perfectly screened layer. Due to the absence of the uncompensated bound charge and external field, both depolarization ($E_n$) and stray ($E_t$) components of the internal electric field are zero inside the flat layer [see the top row in **Fig. 4(a)** and **4(b)**]. The tangential polarization component, $P_t$, which is proportional to $E_t$, is also zero. Normal components of elastic strain and stress tensors, $\sigma_{nn}$ and $u_{nn}$, either appear or change their sign in the region of the FI1-FI2 transition (see the top rows in **Figs. S1**). Other components of elastic strains and stresses are zero inside the flat layer.

With an increase in the layer bending below the critical value, corresponding to the increase in the force density from 0 to 3.7 MPa, the substitution of FI2 regions by FI1 regions occurs near the fixed edges, and the size of the FI2 regions decreases proportionally to the deflection of the layer central part [see two middle rows in **Figs. 3(a)** and **3(b)**]. When the bending exceeds the critical value, corresponding to the critical force density $\sigma_f^{cr} \approx 3.8$ MPa, the FI2 states disappear near the edges, but they simultaneously appear entire the "inflection" regions where the layer is locally flat, as shown in **Fig. 2(c)**. When the bending increases further the FI2 states in the inflection regions expand, and the central region of FI1 states, where the bending and elastic field gradients are maximal, shrinks [see three bottom rows in **Figs. 3(a)** and **3(b)**]. Thus, the critical bending of the layer creates the FI2 states (with $P_n \approx 0.4$ μC/cm$^2$) inside the inflection regions, and significantly increases the normal polarization $P_n$ (up to 4 μC/cm$^2$) in the remainder of the FI1 regions. The wedge-shaped small regions, where the isostructural FI1-FI2 transition occurs, contain the changes of $P_n$ from 2 μC/cm$^2$ to $\pm 0.5$ μC/cm$^2$. The $P_t$ is maximal (about $\pm 0.4$ μC/cm$^2$) at the boundaries of the inflection regions. The physical origin of nonzero $P_t$ is the stray field, $E_t$, which is induced by the uncompensated bound charge in the bended regions. Thus, $P_t$ is proportional to $E_t$, which is in order of magnitude smaller than $E_n$ and has a well-localized maxima at the boundaries of the inflection regions [see the middle and bottom rows in **Fig. 4(a)** and **4(b)**]. Note that $E_n$ is negligibly small inside the inflection regions, because here the CIPS layer has small polarization and is locally flat.

The distribution of elastic strains and stresses changes significantly with increase of bending and starts to deviate strongly from the values for the flat layer once the bending exceeds a certain critical value [compare the middle and bottom rows of **Figs. S1(a)-(f)** in **Appendix B**]. Since the surfaces of the CIPS layer are mechanically free, except for its fixed edges, the stresses $\sigma_{nn}$ and $\sigma_{tn}$, which can reach (50 – 100) MPa, are much smaller than the tangential stress $\sigma_{tt}$, which can reach 1 GPa for the maximal force density ≈10 MPa [see **Fig. S1(d)-(f)**]. At the same time, the strains $u_{nn}$, $u_{nt}$, and $u_{tt}$ are of the same order (≈1% − 1.5%) for the same maximal force density [see **Fig. S1(a)-(c)**]. The normal strain $u_{nn}$ is zero in the inflection regions, as well as near the fixed edges of the layer. The normal stress $\sigma_{nn}$ is very small in the regions of FI1 and FI2 states, and reaches maxima and minima at the FI1-FI2 boundaries. The strain $u_{tn}$ and stress $\sigma_{tn}$ are zero in the flat layer; they gradually increase with the bending increase. For bending more than the critical value, $u_{tn}$ and $\sigma_{tn}$ are maximal near the fixed edges and monotonically decreases to zero towards the layer center. The tangential strain $u_{tt}$ and stress $\sigma_{tt}$ are zero in the flat layer. With the bending increase, $u_{tt}$ and $\sigma_{tt}$ reach maxima and minima near the fixed edges and in the layer center, and are very small in the inflection regions. Noteworthy, the influence of the flexoelectric coupling is small enough for the flexoelectric coefficients $|f_{ijkl}|$ <4 V, which is their physically reasonable range [33].

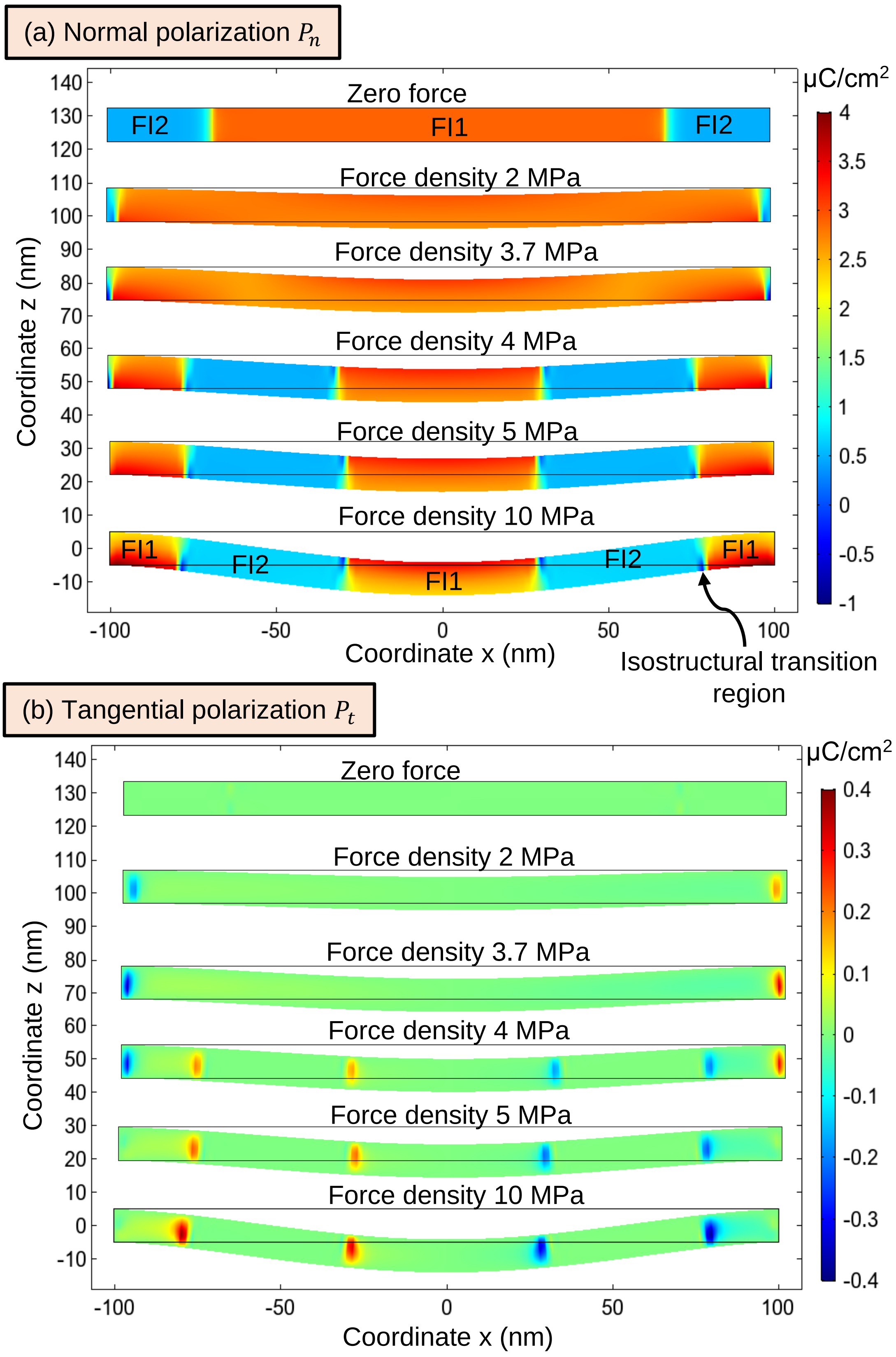

**FIGURE 3**. Distribution of normal and tangential polarization components, $P_n$ **(a)** and $P_t$ **(b)**, inside the CIPS layer with fixed edges (side view) calculated for different magnitudes of applied loading force, which density $\sigma_f$ changes from 0 to 10 MPa (from the top to the bottom rows, respectively). The top and bottom surfaces of the layer are perfectly screened; calculations are performed at 293 K. Color scales show polarization components in µC/cm$^2$. The layer thickness is 10 nm, length is 200 nm, and its width is regarded infinite due to the periodic boundary conditions in the Y-direction.

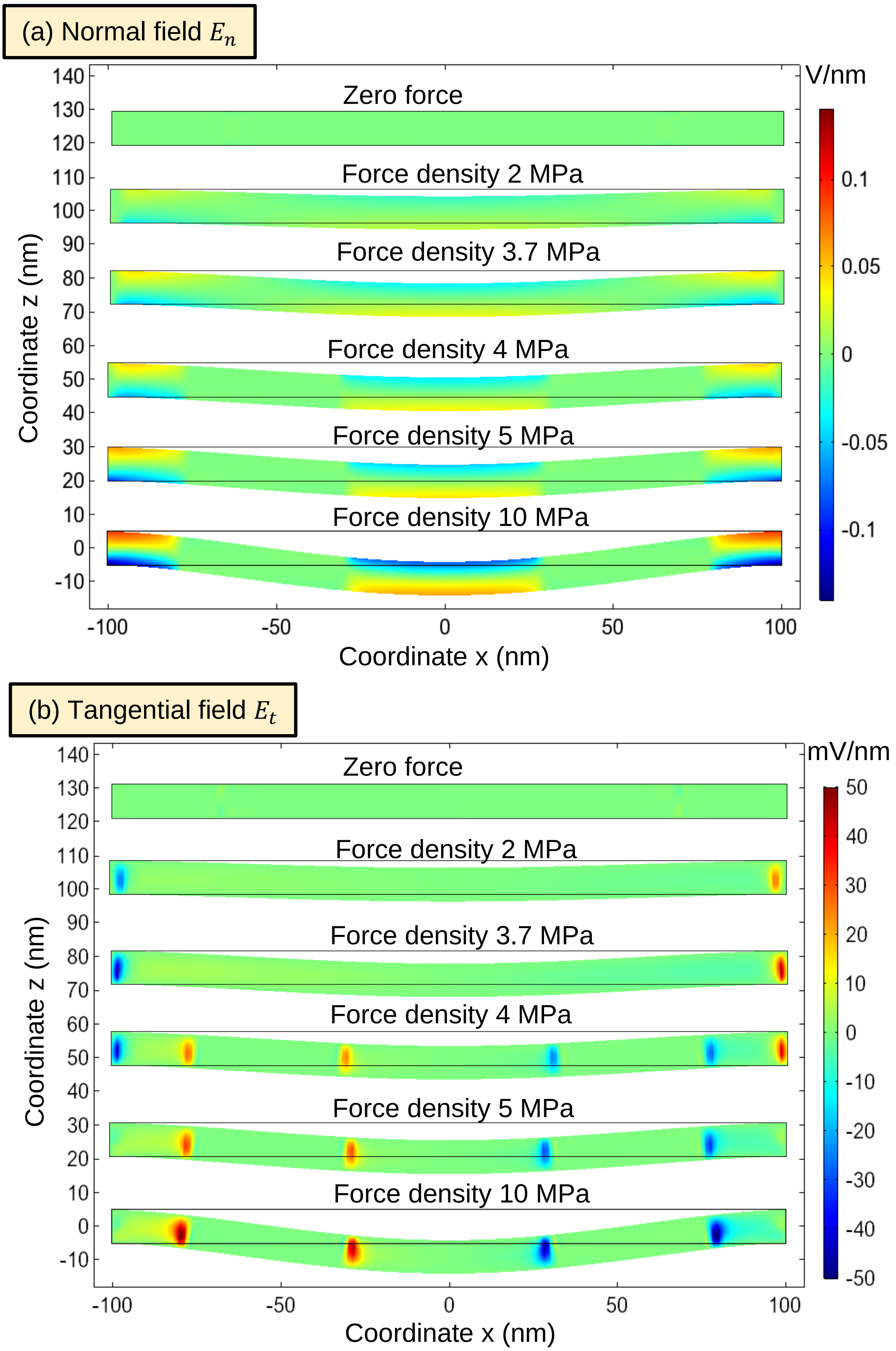

**FIGURE 4**. Distribution of normal and tangential electric field components, $E_n$ **(a)** and $E_t$ **(b)**, inside the CIPS layer with fixed edges (side view) calculated for different magnitudes of applied force, which density $\sigma_f$ changes from zero to 10 MPa (from the top to the bottom rows, respectively). Color scales show the field values in V/nm and mV/nm for the parts **(a)** and **(b)**, respectively. Other conditions are the same as in **Fig. 3.**

The size of FI2 states regions near the fixed edges (orange curve) and in the inflection regions (green curve) of the CIPS layer depends strongly on the magnitude of applied force, as shown in **Fig. 5(a)**. Due to the competition between two FI states, the remainder of the layer is in the FI1 state. Corresponding profiles of normal polarization, $P_n$, calculated for different magnitudes of the force, which density increases from 0 to 1.6 MPa, and then from 1.8 MPa to 6 MPa, are shown in **Fig. 5(b)** and **5(c)**, respectively.

When the force density increases from 0 to 1.5 MPa the edge FI2 regions exist, and their size monotonically decreases from 31 nm to 15 nm, being gradually substituted by the FI1 states. Note that the width of boundary between the edge FI2 states and the central FI1 states is relatively small, and the boundary is almost flat. When the force increases from 1.5 MPa to 2 MPa the size of the edge FI2 states decrease from 15 nm to 5 nm, and then gradually vanish with further increase in the force from 2 MPa to 6 MPa being completely substituted by the FI1 states. The character of the force behavior of the edge FI2 states resembles the "smeared" first order transition.

The FI2 states in the inflection regions far from the layer edges substitute the FI1 states, once the force density exceeds the critical value of about 3.7 MPa. The wedge-shaped FI1-FI2 boundaries in the inflection regions are wider and have a peculiarity (very small or negative polarization) in the four points of maximal curvature. The size of the inflection regions jumps from 0 to 50 nm and then very slowly saturates with further increase in the force density from 3.8 MPa to 10 MPa. Thus, the character of the force behavior of the FI2 states in the inflection regions corresponds to the "sharp" first order transition.

A reasonable question is why the force behavior of the edge FI2 states is similar to the smeared first order transition, and the behavior of the FI2 states in the inflection regions far from the edges is the sharp first order transition? The difference in the transition behavior is related with the difference in the elastic state of the edges and inflection regions. The edges are rigidly clamped from the outer side by external fixing, and the edge FI2 regions induced by the clamping gradually decrease with the bending force increase. The inflection regions, which appear spontaneously, are locally flat and not clamped. The local curvature is zero in the inflection points, and thus the strains $u_{tt}$ and $u_{nn}$ are very small in the inflection regions [see **Fig. S1(a)** and **S1(c)** where]. The four FI1-FI2 boundaries of the inflection regions are self-clamped by the surrounding FI1 states, which are bended and stressed, and thus the normal stress $\sigma_{nn}$ is maximal at the boundaries [see **Fig. S1(e)**].

Noteworthy, the average polarization of the CIPS layer increases with its bending. Microscopically, the enhancement is related with the canting of elementary dipoles (pseudospins) in the bended layer. The result can be confirmed by the DFT calculations.

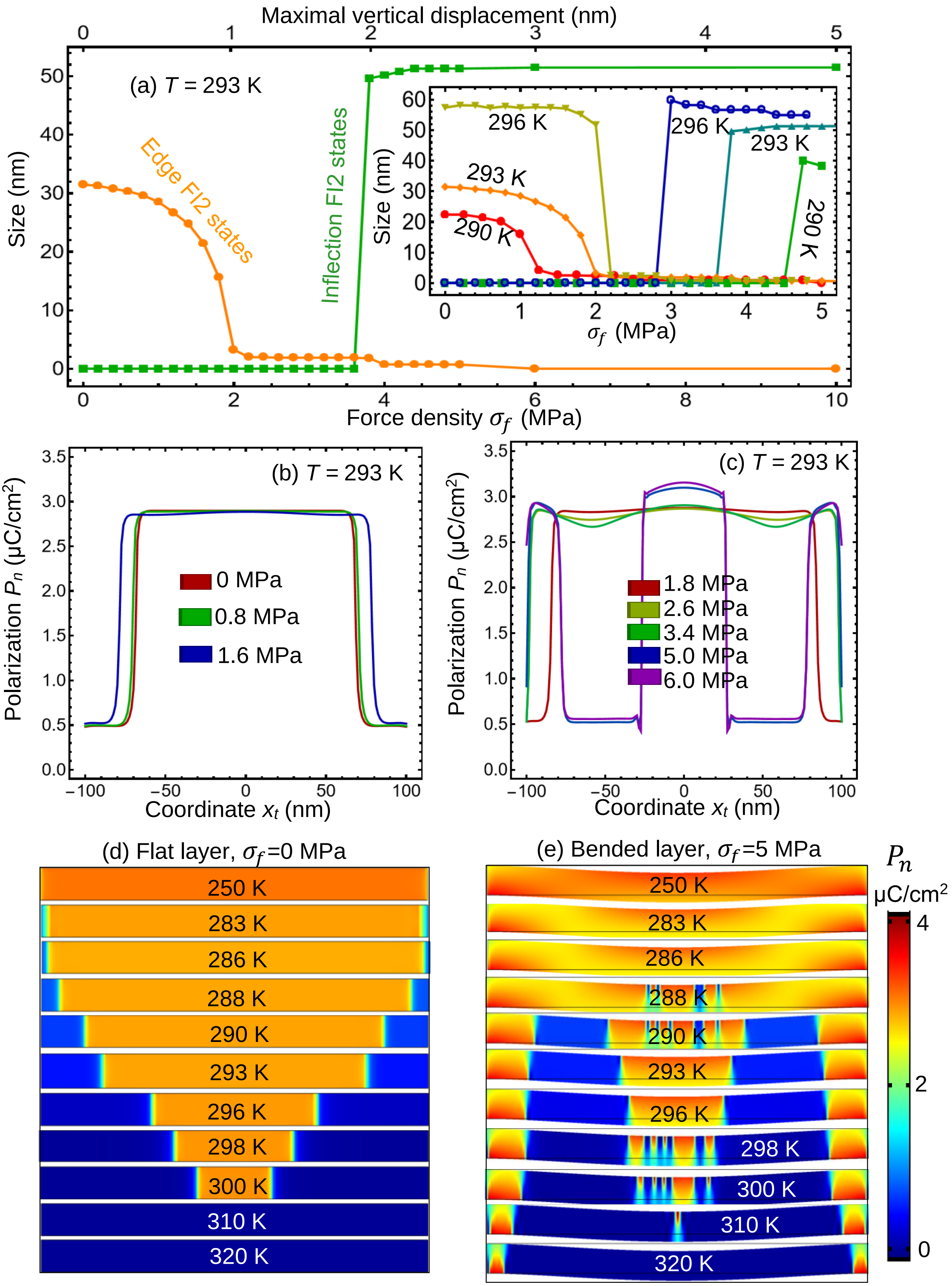

**FIGURE 5**. **(a)** The size of the FI2 states near the fixed edges (orange curve) and in the inflection regions (green curve) via the magnitude of applied force, which density $\sigma_f$ increases from 0 to 10 MPa, $T$ = 293 K. The top scale shows the maximal vertical displacement of the layer. Inset shows the same dependences for 290 K, 293 K and 296 K. **(b-c)** Profiles of normal polarization $P_n$ inside the CIPS layer calculated for different $\sigma_f$, which increases from 0 to 1.6 MPa **(b)**, and then from 1.8 MPa to 6 MPa **(c)**, $T$ = 293 K. **(d-e)** Distributions of normal polarization $P_n$ inside the flat **(d)** and bended **(e)** CIPS layer calculated for different $T$ from 250 K to 320 K (see labels). Other conditions are the same as in **Fig. 3.**

The temperature behavior of the FI2 states near the fixed edges of the flat CIPS layer is characterized by their gradual appearance at 280 K and disappearance at 300 K, which also indicates on the smeared first order transition [see **Fig. 5(d)**]. The temperature behavior of the FI2 states in the inflection regions of the bended CIPS layer is characterized by abrupt appearance at 290 K and disappearance at 300 K, which confirms the sharp first order transition [see **Fig. 5(e)**]. Noteworthy, the isostructural FI1-FI2 transition exists in a relatively narrow temperature range, (280 - 300) K, in the vicinity of CIPS Curie temperature, $T_C$ = 292.67 K.

We have verified that the appearance of FI2 states with small polarization near the fixed edges of the flat or weakly bended CIPS layer, and the appearance of FI2 states in the inflection regions of the strongly bended CIPS layer, as well as the sizes and shape of the FI1-FI2 transition regions, are not related with the shape and/or size of mesh elements used in the FEM. All these unusual effects are specific for CIPS described by the eighth-order LGD functional (1), which also takes into consideration linear and nonlinear electrostriction couplings. Due to the strong and negative nonlinear electrostriction couplings ($Z_{i33} < 0$), and the "inverted" signs of the linear electrostriction coupling ($Q_{33} < 0$, $Q_{23} > 0$ and $Q_{13} > 0$), the bending effect on the polarization re-distribution is complex and unusual for CIPS in comparison with many other ferroelectrics with $Q_{33} > 0$, $Q_{23} < 0$, and $Q_{13} < 0$ (see **Table S1**).

Remarkedly, that the isostructural FI1-FI2 transition and related effects are absent for planar and curved thin layers of other ferroelectrics described by the sixth-order LGD functional, in particular for $Sn_2P_2S_6$, with (or without) consideration of linear and nonlinear electrostriction couplings. Thus, the appearance of the FI2 states and the FI1-FI2 transition, induced by fixing or bending of the thin layer, arises due to the specific four-well potential of the eighth-order LGD functional. Phase diagrams, which contain the FI1 and FI2 states, and the FI1-FI2 transition induced by a rough substrate, are discussed in the next section.

### B. Approximate analytical model of the bended CIPS on a rough substrate

Below we consider a model 2D approximation in Eq.(1) for the normal and tangential strains induced by the layer bending. Using the decoupling approximation for elastic stresses, valid for small bending, nonzero strain components can be estimated as [34]:

$$u_{tt}(x_t, x_n) \approx \frac{2\delta h}{R^4}(x_t^2 + R^2)x_n + Q_{ttij}P_iP_j + Z_{ttij}P_i^2P_j^2, \tag{5a}$$

$$u_{tn}(x_t, x_n) \approx \frac{2\delta h}{R^3}x_t x_n + Q_{tnij}P_iP_j + Z_{tnij}P_i^2P_j^2, \tag{5b}$$

$$u_{nn}(x_t, x_n) \approx \frac{4\nu}{1-\nu}\frac{\delta h}{R^2}x_n + Q_{nnij}P_iP_j + Z_{nnij}P_i^2P_j^2, \tag{5c}$$

where $\nu$ is the Poisson ratio, subscripts $i, j = \{t, n\}$; the local coordinate $x_n$ is counted from the neutral (i.e., unstrained) surface of the layer and varies within the range $-\frac{h}{2} < x_n < \frac{h}{2}$, $\delta h$ is the maximal vertical deflection of the layer, and $R$ is the characteristic radius of curvature of the bended region [**Fig. 2(b)**].

Approximate expressions (5) are in a qualitative agreement with the distributions of elastic strains calculated by FEM and shown in **Fig. 6(a).** It is also seen from **Fig. 6(a)**, that the correlations between the distribution of the normal elastic strain, $u_{nn}$, and polarization components, $P_n$ and $P_t$, are very strong. In particular, the regions with $u_{nn} = 0$, which are inflection regions, coincide with the FI2 states in the spontaneous polarization distribution. Other strains, $u_{tt}$ and $u_{tn}$, are virtually independent on the $P_n$ and $P_t$.

Approximate analytical solutions for the spontaneous polarization are possible if the bending curvature and its characteristic period $K_c$ are much higher than the characteristic length of polarization correlations, $L_c$, in the CIPS layer. The assumption is valid for very smooth corrugations. Here we assume that the corrugation height $h_c$ does not exceed 10 nm, and its characteristic period $K_c$ is about 1 μm. Since $\delta h \le h_c$, $h_c \ll K_c$ and $K_c$ is regarded much higher than $L_c$, and the layer surfaces are perfectly screened by electrodes and/or surface charges, the polarization and strain gradients are relatively small and can be neglected in the first approximation. Thus, one can use the analytical approach of Kvasov and Tagantsev [35] for calculations of the polarization components $P_i$ in dependence on the temperature $T$ and strains $u_{ij}$, at that the role of $u_{tt}$ is qualitatively similar to the influence of mismatch strain in thin films [35]. The approach becomes invalid with the bending increase, because $P_i$ changes the sign [see e.g., **Figs. 3(a)** and **3(b)** for the force density $\sigma_f = 5$ MPa and 10 MPa]. However, we hope that the approach still can be used in the inflection regions, where the strains are small [see **Fig. 6(a)** and **Fig. S1**].

Applying the Legendre transformation of Eq.(1) to the strain-polarization representation, $\tilde{G} = G + u\sigma$, the renormalized free energy density is [11]:

$$\tilde{g}_{LD+ES} = \frac{\alpha^*}{2}P_n^2 + \frac{\beta^*}{4}P_n^4 + \frac{\gamma^*}{6}P_n^6 + \frac{\delta^*}{8}P_n^8 - P_nE_n. \tag{6}$$

The coefficients $\alpha^*$, $\beta^*$, $\gamma^*$, and $\delta^*$ in Eq.(6) depend on the strain $u_{tt}$ as follows:

$$\alpha^* = \alpha + \alpha_1 u_{tt} + \alpha_2 u_{tt}^2, \quad \beta^* = \beta^* + \beta_1 + \beta_2 u_{tt}, \quad \gamma^* = \gamma + \gamma_1 + \gamma_2 u_{tt}, \quad \delta^* = \delta + \delta_1. \tag{7}$$

The dependence of the coefficients $\alpha_1$, $\alpha_2$, $\beta_1$, $\beta_2$, $\gamma_1$, $\gamma_2$ and $\delta_1$ on the electrostriction constants $Q_{ij}$, $W_{ijk}$ and $Z_{ijk}$, and elastic compliances $s_{ij}$ is listed in **Appendix C.**

In Eq.(6), $E_n$ is the sum of external and depolarization fields. Since the electrodes are short-circuited in accordance with the boundary conditions (3b), the external field is assumed to be zero in Eq.(6). The depolarization field can originate the nonzero polarization divergency, $\frac{\partial P_i}{\partial x_i}$, and should accompany the appearance of charged domain walls. For the considered case of perfect electric screening by ideally-conductive electrodes the formation of charged domain walls is very unlikely, and so the depolarization field can be small enough for a flat layer, but it can increase with its bending.

Minimization of the free energy (6) at $E_n = 0$ gives the dependence of the spontaneous polarization, $P_s$, on $T$ and $u_{tt}$, which is shown in **Fig. 6(b)**. The color scale corresponds to the absolute value of $P_s$ in the deepest potential well of the free energy (6). Schematics of a potential relief in the PE phase, and in the ferrielectric states FI1 and FI2 is shown in **Fig. 1(b)**.

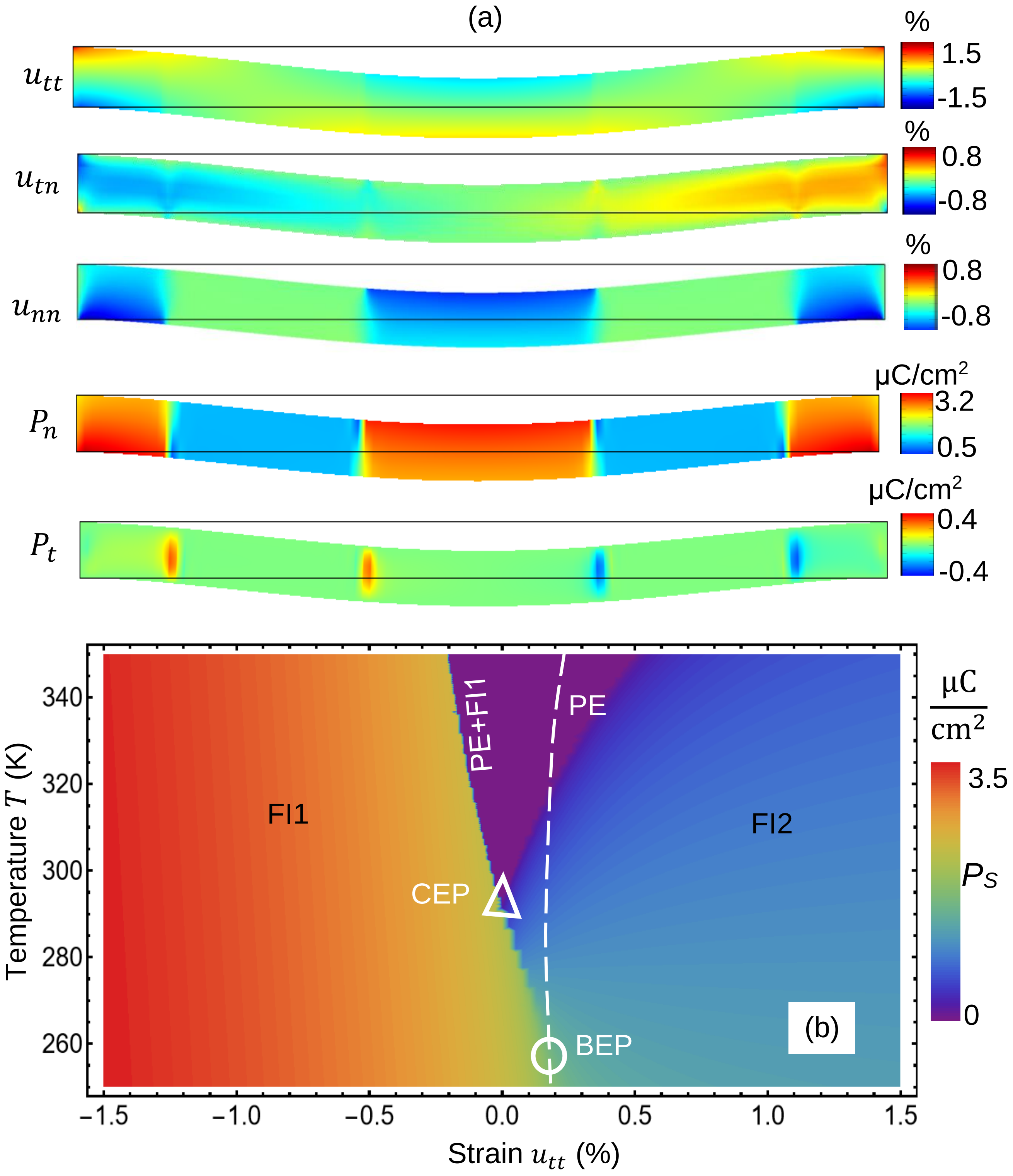

**FIGURE 6**. **(a)** Correlations between the distributions of elastic strains $u_{tt}$, $u_{tn}$ and $u_{nn}$, and polarization components, $P_n$ and $P_t$, in the bended CIPS layer (force density is equal to 5 MPa, $T$ = 293 K). **(b)** The dependence of the spontaneous polarization $P_S$ on temperature $T$ and strain $u_{tt}$. PE is the paraelectric phase, FI1 and FI2 are the ferrielectric states 1 and 2. CEP and BEP are the critical and bicritical end points, marked by the white triangle and circle, respectively. Color scale is the absolute value of $P_s$ in the deepest potential well of the LGD free energy. Other conditions are the same as in **Fig. 3.**

The diagram in **Fig. 6(b)** contains a relatively small triangular-shaped region of the PE phase with $P_s = 0$, which appears at $T \approx 290$ K and expands with the temperature increase. The left part of the diagram is filled by the FI1 state, which has a relatively big magnitude of the spontaneous polarization, $P_s^+$ ≈(1.5 – 3.5) μC/cm$^2$. Since the "sharp" PE-FI1 boundary located at compressive strains $u_{tt} < 0$ and $T > 290$ K is the first order phase transition, the FI1 state can coexist with the PE phase, and the coexistence region is marked as the "FI1+PE". The right part of the diagram is filled by the FI2 state, which has a relatively small magnitude of the spontaneous polarization, $P_s^-$ ≈(0.1 – 1) μC/cm$^2$. The PE-FI2 boundary is the second order phase transition. The diagram also contains the critical end point (**CEP**), where the first order PE – FI phase transition line terminates, and the bicritical end point (**BEP**), where the first order isostructural transition line between the FI1 and FI2 states terminates.

The unusual feature of the diagram in **Fig. 6(b)** is that the high-polarization FI1 state exists at compressive strains $u_{tt} < 0$ and does not vanish for tensile strains $u_{tt} > 0$. Instead, it abruptly (between the CEP and BEP) or continuously (below the BEP) transforms in the FI2 state. Since the magnitude of $P_s$ is big for $u_{tt} < 0$ and small for $u_{tt} > 0$, the situation for the bended CIPS thin layer is untypical for the most strained ferroelectric films, where the out-of-plane polarization is zero or very small for tensile strains, and increases for compressive strains [12, 35].

### C. Bending influence on the ferroelectric polarization reversal

The influence of bending on the ferroelectric polarization reversal in ultrathin CIPS layers is illustrated in **Fig. 7.** Polarization values are averaged over the layer volume, which coincides with the free charge density of the conducting electrodes.

Quasi-static polarization hysteresis loops, shown in **Fig. 7(a)**, are calculated for zero and small magnitudes of applied force density, 0 MPa, 1 MPa and 2 MPa, respectively. For the force densities smaller than 2 MPa the edge FI2 states exist, and the coercive voltage, $V_c$, very slightly increases with increase of the force density, but the remanent polarization, $P_r$, is the same, and the loop shape is rectangular-like and symmetric with respect to the voltage axis [compare the width, height and shape of the black, blue and green loops in **Fig. 7(a)**].

For the force densities higher than 3 MPa the edge FI2 states disappear, but they appear in the inflection regions far from the edges. Due to this, the coercive voltage and the remanent polarization decreases strongly with increase in the force density from 5 MPa to 10 MPa, the loop shape becomes step-like and tilted with respect to the voltage axis [compare the width, height and shape of the dark-yellow, orange, red and dark-red loops in **Fig. 7(b)**]. The steps on the loops originated from the domain structure re-building, and in particular with the "stubborn" domain related with the layer bending (see **Fig. S2,** which shows the distribution of $P_n$ inside the CIPS layer for different applied voltages after several cycles of the periodic voltage application). The coercive voltage decrease is in agreement with the experimental result [17]. Thus, the revealed isostructural FI1-FI2 transition can change significantly the coercive voltage, remanent polarization and the hysteresis loop shape of the ferroelectric polarization reversal in ultrathin CIPS nanoflakes, which is highly required for their advanced applications.

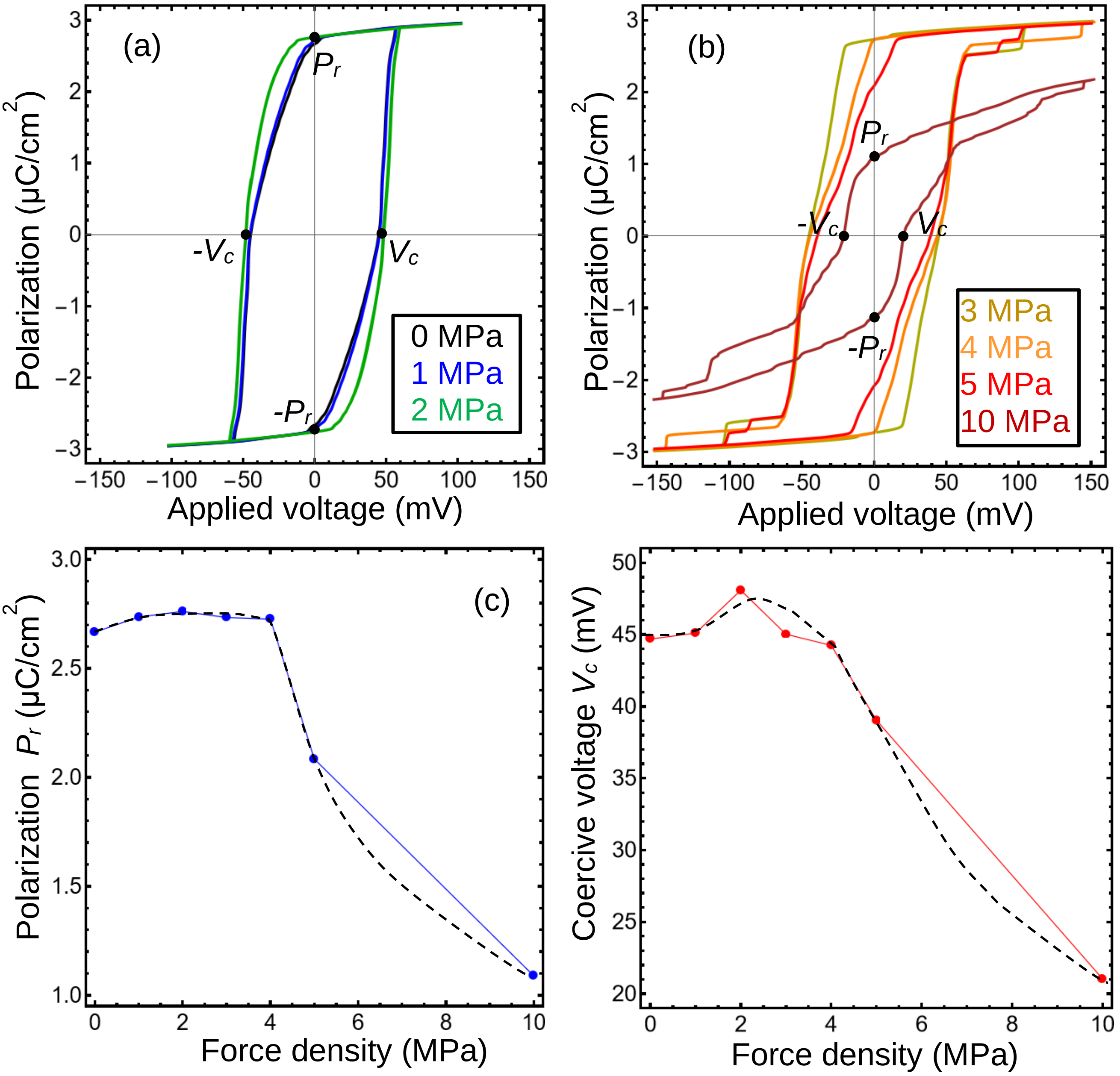

**FIGURE 7. (a)-(b)** Quasi-static hysteresis loops of heaveraged polarization reversal in ultrathin CIPS layer calculated for different magnitudes of applied force density, 0 MPa, 1 MPa and 2 MPa [black, blue and green

loops in the plot **(a)**], 3 MPa, 4 MPa, 5 MPa and 10 MPa [dark-yellow, orange, red and dark-red loops in the plot **(b)**], and $T = 293$ K. **(c)-(d)** The dependence of the coercive voltages **(c)** and remanent polarizations **(d)** on the applied force. Blue and red points are calculated by FEM, dashed black curves are interpolations. Other conditions are the same as in **Fig. 3.**

## IV. CONCLUSIONS

Using the LGD eighth-order thermodynamic potential we analyze the bending-induced re-distribution of electric polarization and field, elastic stresses and strains inside ultra-thin layers of vdW ferrielectrics. As an example, we consider CIPS thin layer with fixed edges and suspended central part, which bending is controlled by external forces, such as attraction force by a rough substrate.

When the top and bottom surfaces of the layer are perfectly screened, the single-domain ferrielectric state FI1 with the out-of-plane polarization $P_n \approx (3 - 5)$ μC/cm$^2$ is stable in the central part of the flat layer, and the FI2 states with an order of magnitude lower polarization, $P_n \approx (0.3 - 0.5)$ μC/cm$^2$, are stable near the fixed edges. With an increase of the layer bending below the critical value, two processes occur. The first process is the substitution of the edge FI2 states by the FI1 states. The second process arises when the bending exceeds the critical value, and the FI2 states appear abruptly in the inflection regions, and expand when the bending increases further.

The behavior of the FI2 states near the fixed edges and in the inflection regions depends strongly on the magnitude of applied force, which causes the bending of the CIPS layer. When the force density increases below the critical value, the size of the FI2 regions near the fixed edges monotonically decreases; and the inflection regions are absent. When the force density increases above the critical value, the size of the edge FI2 regions becomes very small and they are gradually substituted by the FI1 states with further increase of the force. Thus, the character of the force behavior of the FI2 states near the fixed edges resembles the smeared first order transition. The FI2 states in the inflection regions appear spontaneously once the force density exceeds the critical value, and their size very slowly saturates with further increase of the force. The character of the force behavior of the FI2 states in the inflection regions corresponds to the sharp first order transition. The temperature behavior of the FI2 states near the fixed edges also indicates on the smeared first order transition, and the temperature behavior of the FI2 states in the inflection regions of the bended CIPS layer confirms the step-like first order transition. The isostructural FI1-FI2 transition exists in a relatively narrow temperature range, (280 - 300) K, in the vicinity of CIPS Curie temperature, $T_C$ = 292.67 K.

The isostructural FI1-FI2 transition is specific for those bended vdW ferrielectrics, which are described by the eighth (or higher) order LGD functional with consideration of linear and nonlinear electrostriction couplings. Namely, the FI1 and FI2 states, induced by bending of ultrathin CIPS layers, arise due to the specific four-well potential of the eighth-order LGD functional. The calculated phase diagrams show that the transition between the FI1 and FI2 states can be induced by the tangential strains created by the rough substrate.

The revealed isostructural FI1-FI2 transition can significantly reduce the coercive voltage of ferroelectric polarization switching in ultrathin CIPS nanoflakes, which is urgent for their advanced applications.

**Acknowledgements.** S.V.K., Y.L., and J.Y. are supported by the center for 3D Ferroelectric Microelectronics (3DFeM), an Energy Frontier Research Center funded by the U.S. Department of Energy (DOE), Office of Science, Basic Energy Sciences under Award Number DE-SC0021118. A.N.M. acknowledges support from the National Research Fund of Ukraine (project "Low-dimensional graphene-like transition metal dichalcogenides with controllable polar and electronic properties for advanced nanoelectronics and biomedical applications", grant application 2020.02/0027). E.A.E. and N.V.M. acknowledge support from the National Academy of Sciences of Ukraine. A.L.K. and A.N.M. also acknowledge support from the European Union's Horizon 2020 research and innovation programme under the Marie Skłodowska-Curie grant agreement No 778070. This work (A.L.K.) was developed within the scope of project CICECO-Aveiro Institute of Materials (UIDB/50011/2020 & UIDP/50011/2020) financed by national funds through the FCT—Foundation for Science and Technology (Portugal).

**Data availability statement.** Numerical results presented in the work are obtained and visualized using a specialized software, Mathematica 13.2 [36]. The Mathematica notebook, which contain the codes, is available per reasonable request.

**Authors' contribution.** The research idea belongs to S.V.K. A.N.M. formulated the problem, performed analytical calculations, analyzed results, and wrote the manuscript draft. E.A.E. wrote codes. Y.M.V., N.V.M. and S.V.K. worked on the results explanation and manuscript improvement. All co-authors discussed the obtained results.

# Supplementary Materials

## APPENDIX A. LGD parameters for a bulk ferroelectric $CuInP_2S_6$

The values $T_C$, $\alpha_T$, β, $\gamma$ and $\delta$, $Q_{ijkl}$ and $Z_{ijkl}$ in **Table SI** were defined for CIPS in Ref.[37] from the fitting of experimentally observed temperature dependence of dielectric permittivity [38, 39, 40], spontaneous polarization [41] and lattice constants [42] for hydrostatic and uniaxial pressures. The elastic compliances $s_{ij}$ were estimated from the ultrasound velocity measurements [43, 44, 45].

**Table SI.** LGD parameters for a bulk ferroelectric $CuInP_2S_6$ at fixed stress

| **Coefficient** | **Units** | **Numerical value** |
|---|---|---|
| $\varepsilon_b \approx \varepsilon_s$ | dimensionless | 9 |
| $\alpha_T$ | $C^{-2}\cdot m$ J/K | $1.64067\times10^{7}$ |
| $T_C$ | K | 292.67 |
| $\beta$ | $C^{-4}\cdot m^5 J$ | $3.148\times10^{12}$ |
| $\gamma$ | $C^{-6}\cdot m^9 J$ | $-1.0776\times10^{16}$ |
| $\delta$ | $C^{-8}\cdot m^{13} J$ | $7.6318\times10^{18}$ |
| $Q_{i3}$ | $C^{-2}\cdot m^4$ | $Q_{13} = 1.70136 - 0.00363\,T$, $Q_{23} = 1.13424 - 0.00242\,T$, $Q_{33} = -5.622 + 0.0105\,T$ |
| $Z_{i33}$ | $C^{-4}\cdot m^8$ | $Z_{133} = -2059.65 + 0.8\,T$, $Z_{233} = -1211.26 + 0.45\,T$, $Z_{333} = 1381.37 - 12\,T$ |
| $W_{ij3}$ | $C^{-2}\cdot m^4$ $Pa^{-1}$ | $W_{113} \approx W_{223} \approx W_{333} \cong -2\times10^{-12}$ |
| $s_{ij}$ | $Pa^{-1}$ | $s_{11} = 1.510\times10^{-11}$ , $s_{12} = 0.183\times10^{-11}$ |
| $g_{33ij}$ | J $m^3/C^2$ | $g \cong 2\times10^{-9}$ |

**APPENDIX B. Finite element modelling for elastic strains, stresses and polarization reversal in the bended CIPS layer**

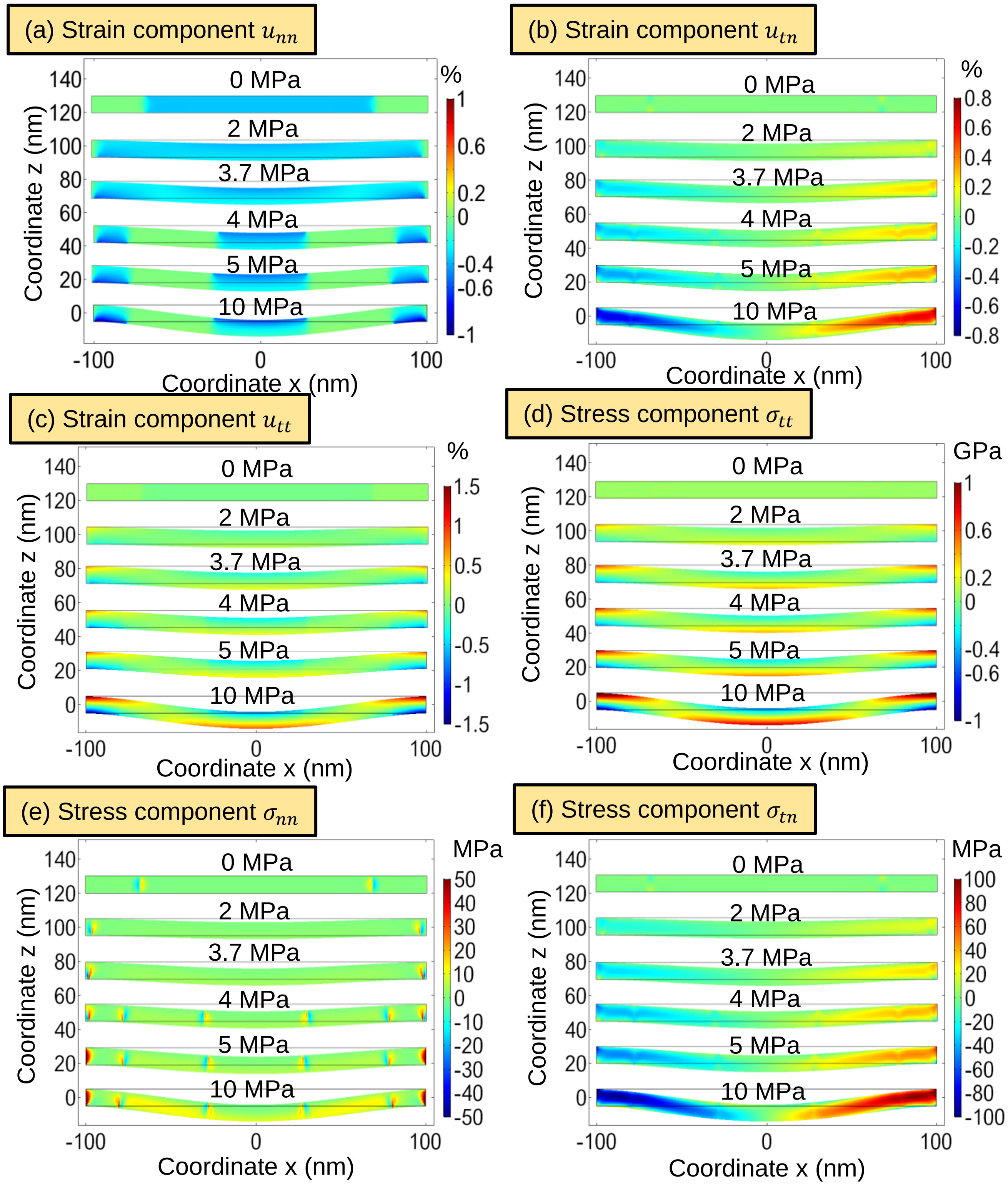

**FIGURE S1**. Distribution of elastic strain tensor components, $u_{nn}$ **(a)**, $u_{nt}$ **(b)**, and $u_{tt}$ **(c)**; and stress tensor components, $\sigma_{tt}$ **(d)**, $\sigma_{nn}$ **(e)**, and $\sigma_{tn}$ **(f)** inside the 10-nm thick CIPS layer with fixed edges (side view), calculated for different magnitudes of applied force, which density changes from 0 to 10 MPa (from the top to the bottom rows, respectively). Color scales show the strain values in percents for the parts **(a)-(c)**, and in GPa and MPa for the parts **(d)-(f)**, respectively. Other conditions are the same as in **Fig. 3.**

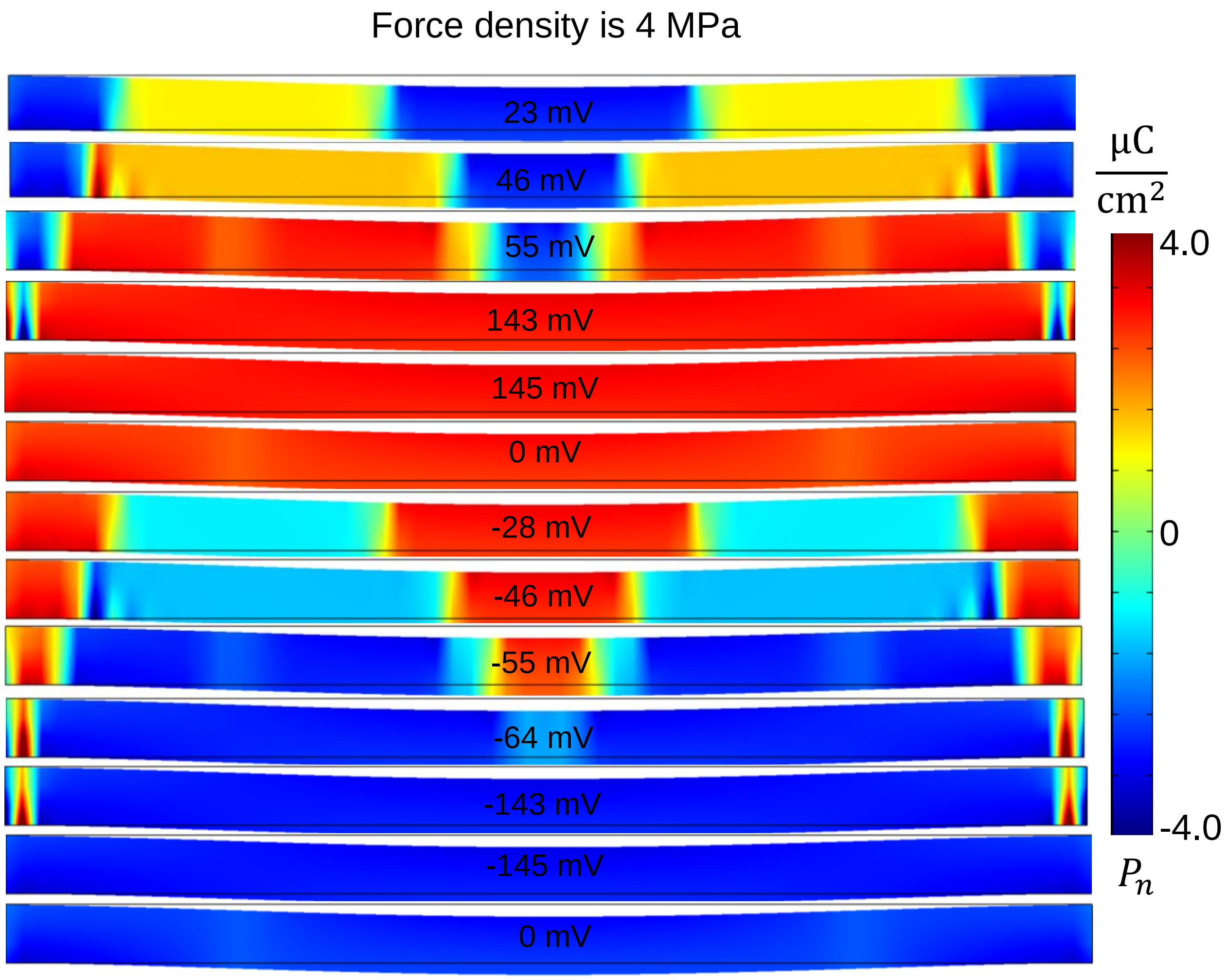

**FIGURE S2**. Distribution of $P_n$ inside the CIPS layer calculated for different applied voltages after multiple cycles of the periodic voltage application (see legends). The force density is 4 MPa. Other conditions are the same as in **Fig. 3.**

From **Fig. 7(d)** and **S2**, the homogeneous field as high as (50 - 20) mV/nm is required to reverse the spontaneous polarization of the CIPS layer. From **Fig. 4(a)**, the sign-alternating field $E_n$ as high as 100 mV/nm exists in the 10-nm wide FI1 regions. However, the high local sign-alternating field appeared not enough for the local polarization reversal, because its average value across the 10-nm wide regions is very small or rigorously zero. To reverse the polarization, $E_n$ should overcome the thermodynamic coercive field in a definite spatial region, and the higher is the field the smaller can be the size of the field-averaging region, however it should not be smaller than the correlation volume.

The CIPS layer transforms into another domain state after multiple cycles of periodic voltage application (as shown in **Fig. S2**), where the polarization in the FI1 and FI2 states has different signs, i.e., the layer splits into FI1 and FI2 domains. The poly-domain configuration creates lower depolarization field, as expected.

## APPENDIX C. Free energy with renormalized coefficients

Using results [35] and making the Legendre transformation of Eq.(1) to the strain-polarization representation, $\tilde{G} = G + u\sigma$, the renormalized free energy density is [46]:

$$\tilde{g}_{LD+ES} = \frac{\alpha^*}{2}P_3^2 + \frac{\beta^*}{4}P_3^4 + \frac{\gamma^*}{6}P_3^6 + \frac{\delta^*}{8}P_3^8 - P_3E_3. \tag{S.1}$$

The coefficients $\alpha^*$, $\beta^*$, $\gamma^*$, and $\delta^*$ in Eq.(5) depend on the strain, $u_m$, as:

$$\frac{\alpha^*}{2} = \frac{\alpha}{2} + \frac{u_m(Q_{23}(s_{12}-s_{11})+Q_{13}(s_{12}-s_{22}))}{s_{11}s_{22}-s_{12}^2} - u_m^2\frac{(s_{12}-s_{22})^2W_{113}+(s_{11}-s_{12})^2W_{223}}{2(s_{12}^2-s_{11}s_{22})^2}, \tag{S.2a}$$

$$\frac{\beta^*}{4} = \frac{\beta}{4} + \frac{Q_{23}^2s_{11}-2Q_{13}Q_{23}s_{12}+Q_{13}^2s_{22}}{2(s_{11}s_{22}-s_{12}^2)} + u_m\left\{\frac{(s_{22}-s_{12})Z_{133}+(s_{11}-s_{12})Z_{233}}{s_{12}^2-s_{11}s_{22}} + \frac{Q_{23}(s_{12}(s_{12}-s_{22})W_{113}+s_{11}(s_{11}-s_{12})W_{223})+Q_{13}(s_{22}(-s_{12}+s_{22})W_{113}+s_{12}(-s_{11}+s_{12})W_{223})}{(s_{12}^2-s_{11}s_{22})^2}\right\}, \tag{S.2b}$$

$$\frac{\gamma^*}{6} = \frac{\gamma}{6} + \frac{(Q_{13}s_{22}-Q_{23}s_{12})Z_{133}+(Q_{23}s_{11}-Q_{13}s_{12})Z_{233}}{s_{11}s_{22}-s_{12}^2} - \frac{-2Q_{13}Q_{23}s_{12}(s_{22}W_{113}+s_{11}W_{223})+Q_{23}^2(s_{12}^2W_{113}+s_{11}^2W_{223})+Q_{13}^2(s_{22}^2W_{113}+s_{12}^2W_{223})}{2(s_{12}^2-s_{11}s_{22})^2} + u_m\frac{s_{22}^2W_{113}Z_{133}+s_{11}^2W_{223}Z_{233}-s_{12}(s_{22}W_{113}+s_{11}W_{223})(Z_{133}+Z_{233})+s_{12}^2(W_{223}Z_{133}+W_{113}Z_{233})}{(s_{12}^2-s_{11}s_{22})^2}, \tag{S.2c}$$

$$\frac{\delta^*}{8} = \frac{\delta}{8} + \frac{s_{22}Z_{133}^2-2s_{12}Z_{133}Z_{233}+s_{11}Z_{233}^2}{2(s_{11}s_{22}-s_{12}^2)} + \frac{Q_{23}(s_{12}(s_{22}W_{113}+s_{11}W_{223})Z_{133}-s_{12}^2W_{113}Z_{233}-s_{11}^2W_{223}Z_{233})}{(s_{12}^2-s_{11}s_{22})^2} + \frac{Q_{13}(-s_{22}^2W_{113}Z_{133}+s_{12}s_{22}W_{113}Z_{233}+s_{12}W_{223}(-s_{12}Z_{133}+s_{11}Z_{233}))}{(s_{12}^2-s_{11}s_{22})^2}. \tag{S.2d}$$

Since $W_{ijk}$ are small enough, we remain only linear terms in $W_{ijk}$ in Eqs.(S.2), and omit all terms proportional to higher powers of the parameter. The higher powers of $W_{ijk}$ lead to the 12-th powers of polarization in the renormalized free energy density (S.1), and a general case of 12-th power potential is considered in Ref.[47]. We account for these cumbersome and small higher renormalization terms numerically.